# Spectroscopic observation of planetary nebulae

**Paul Luckas**

This paper describes how commercially available spectrographs can be used to identify and measure some basic characteristics of planetary nebulae.

## Introduction

Planetary nebulae present interesting targets for small telescope spectroscopy for a number of reasons. For the relative newcomer, their observation presents a logical 'next step' in commissioning a new spectrograph, or progressing beyond the basics of stellar classification. Displaying prominent emission lines of ionised hydrogen and doubly ionised oxygen (common features of emission nebulae), 'planetaries' provide excellent spectroscopic examples of low density gas ionisation by a very hot source, a recurring theme in much of stellar astrophysics. As such, they vividly demonstrate the second of Kirchhoff's laws of spectroscopy.

Gustav Kirchhoff, a mid-19th century German physicist, was the first to conceptualise the three observable phenomena of continuous, emission and absorption spectra.[1] In emission spectra, low pressure gas which has been excited produces emissions at specific wavelengths according to the elements present in the gas. In a neon lamp, the source is an electric current. In planetary nebulae, the source of this 'excitation' is an extremely hot central star, a white dwarf with surface temperatures exceeding 30,000 K. Planetary nebulae occupy a key stage in the evolution of low- and intermediate-mass main sequence stars beyond the red giant phase. Our Sun is one such star, and will quite possibly produce a planetary nebula in its retirement some 4 to 5 billion years from now.

One of the first astronomers to investigate planetary nebulae spectroscopically was William Huggins in the 19th century. Huggins discovered that these objects were gaseous, rather than stellar in nature, through the observation of bright emission lines rather than the continuous absorption spectra found in stars at the time. Since then, planetary nebulae have been the subject of ongoing research, much of it focused on the nature of their central stars and on atomic morphologies and motions across various points in their extended nebulae.[2] There is little doubt that the spectroscopic investigation of planetary nebulae presents a viable and continuing area of study for amateur astronomers.

This paper provides examples of both low- and high-resolution spectra, identifying and measuring some basic characteristics of these fascinating targets. The telescope used is a relatively modest-sized 0.35m Ritchey–Chrétien (RC) design, situated in a low altitude urban environment under the light polluted skies of Perth, Western Australia.

## Instrumentation

The 0.35m RC telescope at Shenton Park Observatory utilises a multiple instrument package image train consisting of a photometric camera and both high and low resolution spectrographs, individually selectable via an optical manifold. The *Shelyak Alpy 600* low-resolution spectrograph[3] consists of a dispersing element comprising a transmission grating + prism (grism) with a density of 600 lines/mm. A guide module utilising a reflective slit and a calibration module containing a remotely switchable Ar/Ne gas discharge lamp complete the assembly. The typical spectral dispersion is 480Å/mm and a spectral resolution of R≈ 540 at 5852Å is achieved in normal practice. A full first order visible spectrum is contained within the sensor area of an attached *Atik 414EX* CCD camera.[4]

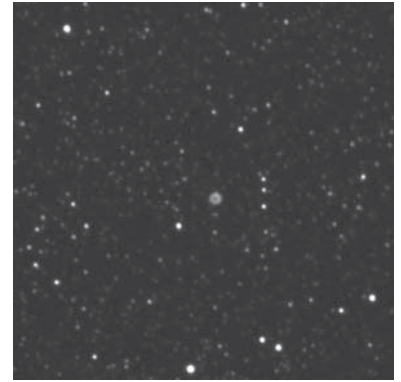

**Figure 1.** Many planetary nebulae appear as small and indistinct objects. This 60 second exposure, taken by the author, shows NGC 6565 within a 5 arcminute field of view as an object little more than 10 arcseconds across.

The *Shelyak Lhires III*[3] is a Littrow design, high resolution spectrograph fitted with a 2400 lines/mm reflection grating. It incorporates the same Ar/Ne and tungsten calibration system used in the *Alpy*. The *Lhires* produces a typical dispersion of 15Å/mm and a spectral resolution of R≈ 16,000 at 6532Å. A typically narrow range of approximately 120Å is displayed on the sensor of the attached *Atik 314L+* CCD camera[4] at this target wavelength.

Attached to the third port is an *SBIG ST-10 (KAF-3200)* camera[5] equipped with Clear, B and V filters used primarily for photometry.

An independent guide camera is integrated into the telescope's optical manifold via a pick-off mirror. As a beneficial consequence, the guide ports on both spectrographs have been repurposed with video cameras to allow for live target centring on the slit. This instantaneous feedback facilitates very precise target/slit positioning of bright stellar objects, but the poor sensitivity of the video cameras makes it difficult to accurately position a faint central star or a particular area of interest on the slit when targeting planetary nebulae. The live view of a planetary is characteristically that of a noisy small faint fuzzy 'blob' and as such, is positioned on the slit 'centrally' in as best a fashion as can be achieved.

## Test images

Three planetary nebulae (among a dozen or so) were chosen at random, situated overhead near the meridian at the time of observation in late 2017 July. Aside from visibility, the only other consideration was magnitude – all three should be within a visual magnitude limit of 10. Owing to the time of year, test images were taken under poor skies including cloud. At best, only one or two short exposures were possible with the *Alpy*, combined into a total integration time of no more than 5 minutes per target. Moreover, the onset of rain prevented the acquisition of reference star spectra for atmospheric and instrumental response correction. All spectral images were bias, dark and flat field corrected, with the latter accomplished using a tungsten lamp integrated into the calibration module of the *Alpy*.

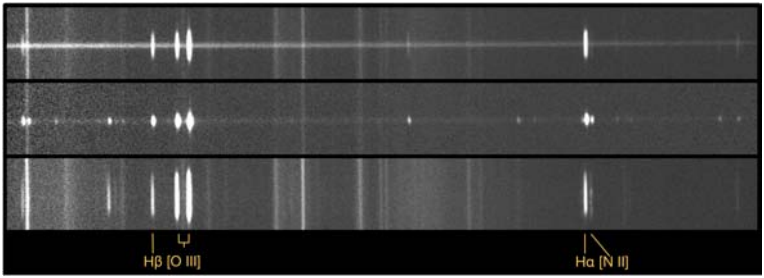

**Figure 2.** Raw test spectral images of NGC 6629, NGC 6644 and NGC 6818. Square brackets are conventionally used when identifying 'forbidden' lines in spectra.

## Test results

Figure 2 shows test exposures for (top to bottom) NGC 6629, NGC 6644 and NGC 6818. The wavelength range has been cropped to show the visible spectrum spanning from 4200Å at left to 7200Å at right. Even in this form and at this low resolution, the spectral images reveal a number of interesting features:

The 'heights' of the bright spectral emission lines show the comparative sizes of the three nebulae as vertical extensions along the spectrograph's slit, with the middle image showing NGC 6644 to be that of a more 'compact' planetary nebula.

The characteristic [OIII] 4959Å and 5007Å emission doublet features prominently a quarter way in from left on each of the three targets. These are the famous 'nebulium' lines; the forbidden transitions of doubly ionised oxygen at very low density first observed in nebulae in the 19th century and proposed at the time as the discovery of a new element. The bright emission immediately left of the [O III] doublet is the hydrogen Balmer Hβ line.

The hydrogen Balmer Hα line situated a quarter way in from the right includes a typical and faint adjacent [N II] 6584Å emission in the bottom two spectra. Subtle changes in the brightness along the length of the [N II] emission in NGC 6818 (bottom image) hint at the varying complexity across the extended regions of this more diffuse nebula.

NGC 6629 at top shows an evident and comparatively bright continuum (the horizontal band). An apparent visual brightening towards the shorter wavelengths at left would appear as evidence for a 'hot' white dwarf central star emitting most of its radiation in the ultraviolet. However, without proper instrument response correction and at this low resolution, there may be reason to suspect contributions from other sources, such as the compact central region, rather than the magnitude 13.3 star alone. It should be noted that the central stars of most planetary nebulae are visually very faint and in many cases no central stars have been observed or are at least spectroscopically 'contaminated' by the surrounding gas. In any event, the inability to guarantee that the central star was positioned accurately in the slit precludes comparison of this phenomena across test images.

The common, broad and unfortunately plentiful emissions that extend the full vertical height of all images are background light pollution emissions. These are removed during processing as part of software binning and sky removal algorithms, to produce near perfect spectral profiles of the target.

Notwithstanding the rudimentary appearance of the test images, spectra such as these parallel the foundational work of early 20th century astronomers using photographic plate technology and analogue analysis techniques. In spite of the software processing pipelines that automate much of today's digital image reduction, basic spectral images still serve to show the fundamentals of the dispersion of light, and its revelations of atomic theory present in exotic and distant astronomical phenomena. We are, quite literally, observing Kirchhoff's second law in action!

## Data acquisition & image processing

An additional 'faint' target, NGC 6565, was added to the list of planetary nebulae subsequently imaged on the clear nights of 2017 August 1 & 2. With good weather, it was possible to accumulate integrations of up to 40 minutes (using 10 minute sub exposures) together with calibration spectra. Additional spectra of the star HD 175892, a type A1V star, were acquired for instrumental and atmospheric response correction.

While bias, dark and flat field correction was facilitated using *MaxIM DL*[6] as part of the image acquisition pipeline, the bulk of spectral processing was accomplished using *ISIS*.[7] A wavelength calibration solution for the *Alpy* was pre-determined using the 'calibration assistant' in *ISIS* which in itself utilises a combination of lines from the *Alpy*'s internal Ar/Ne lamp and the hydrogen Balmer

### Table 1. Target information

| PN | RA (2000) | Dec (2000) | V mag | Total exposure(s) |
|---|---|---|---|---|
| NGC 6565 | 18 11 52 | −28 10 42 | 12.2 | 2400 |
| NGC 6629 | 18 25 42 | −23 12 10 | 9.9 | 1800 |
| NGC 6644 | 18 32 34 | −25 07 44 | 9.6 | 900 |
| NGC 6818 | 19 43 58 | −14 09 13 | 9.3 | 600 |

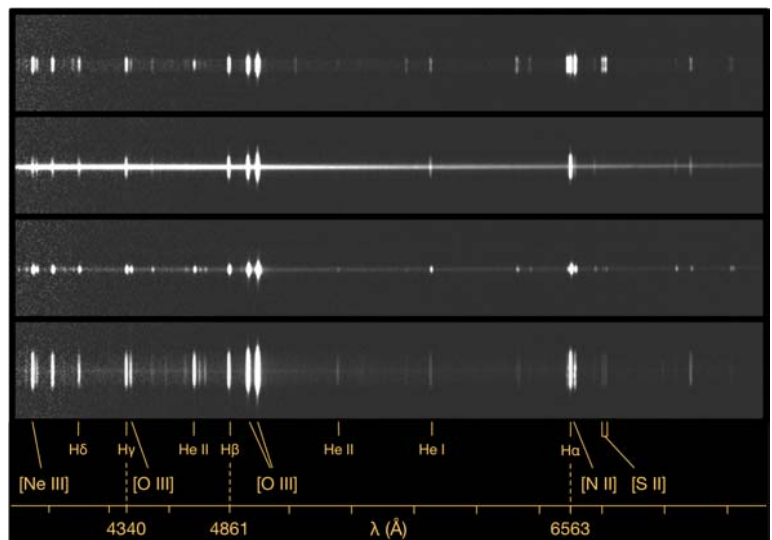

**Figure 3.** *Top to bottom:* Fully processed spectral images of NGC 6565, NGC 6629, NGC 6644 and NGC 6818.

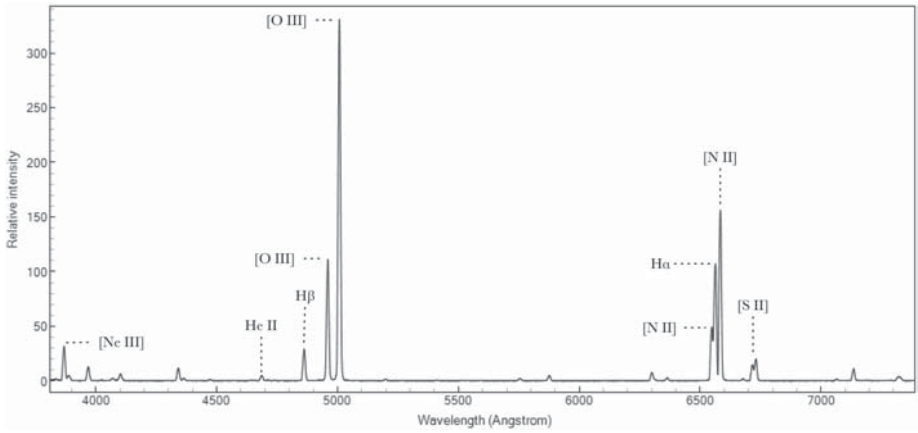

Figure 4. The basic low resolution, software-produced spectral profile of NGC 6565.

tance above and below the spectrum serve as satisfactory guides for processing the spectra of planetary nebulae. *ISIS* employs sophisticated algorithms during processing to combine these and various other routines in the automated creation of a resulting one-dimensional (1D) spectral profile.

Figure 3 shows the target spectral images fully processed in *ISIS*. Individual sub-exposures have been co-added and geometric correction has been applied. Note the absence of light pollution emissions, now completely subtracted as part of software processing.

Longer integration times under better seeing resulted in significantly better signal to noise ratios compared to the test images shown in Figure 2. Subtle variations in the [N II] 6583Å emission of NGC 6818 (bottom image) are more pronounced and the [N II] 6548Å emission line is now distinguishable on the shorter wavelength (left) side of Hα. This 'stratification' in the [N II] pair is particularly evident in the low-excitation spectral lines of diffuse planetaries and easily detected at the *Alpy's* resolution. Also now obvious, and typical of the high ionisation levels of planetary nebulae, is the presence of recombination lines of ionised helium, notably the bright He II 4686Å emission left of Hβ.

absorption features of an A-type star also imaged with the *Alpy*. This combined method alleviates the lack of strong emission lines at shorter wavelengths using the calibration lamp alone, though care must be taken to ensure that the radial velocity (manifesting itself as Doppler shift in the absorption lines) of the A-type star is small compared to the desired calibration accuracy. The resulting wavelength solution has proven robust in tests of standard stars[8,9] and the *Alpy* itself has shown to be extremely stable under varying seasonal conditions and nightly temperature fluctuations. The horizontal pixel location of the Ne 5852Å line as measured in *ISIS* during wavelength calibration has not shifted over the lifetime of the current installed optical train (over one year). Vertical position of the spectrum is similarly stable to within a pixel.

Creation of an instrumental response and atmospheric correction profile is accomplished using spectra taken of a nearby A-type star, processed again in *ISIS* using its integrated *Miles* database[10] for reference standards and response curve creation tools. Geometric corrections of spectral image 'tilt' and calibration line 'smile' are accomplished similarly within the *ISIS* interface, as are the definitions for spectrum binning and sky subtraction zones. A binning size encompassing the vertical height of prominent emission features and sky background zones of roughly the same size defined some dis-

## Spectral profiles

While the improved spectral images reveal meaningful information on temperature and composition, converting these into 1D spectral profiles allows for more quantitative analysis. Figure 4 shows the *ISIS*-produced spectral profile of NGC 6565 as displayed in *PlotSpectra*,[11] illustrating a typical 1D low resolution profile of a planetary nebula plotted along a wavelength scale. The profile has been scaled to unity at around 6200Å, allowing the relative flux intensities and ratios of various emission features to be more accu-rately visualised. The [O III] emission doublet centred around 5000Å together with Hα at 6563Å and its enclosing [N II] 6548Å and 6583Å lines still dominate the spectrum, though we can now better visualise the relative strengths of these lines, with [O III] 5007Å clearly the more intense emission. Also visible are additional forbidden [S II] lines at 6716Å and 6730Å respectively.

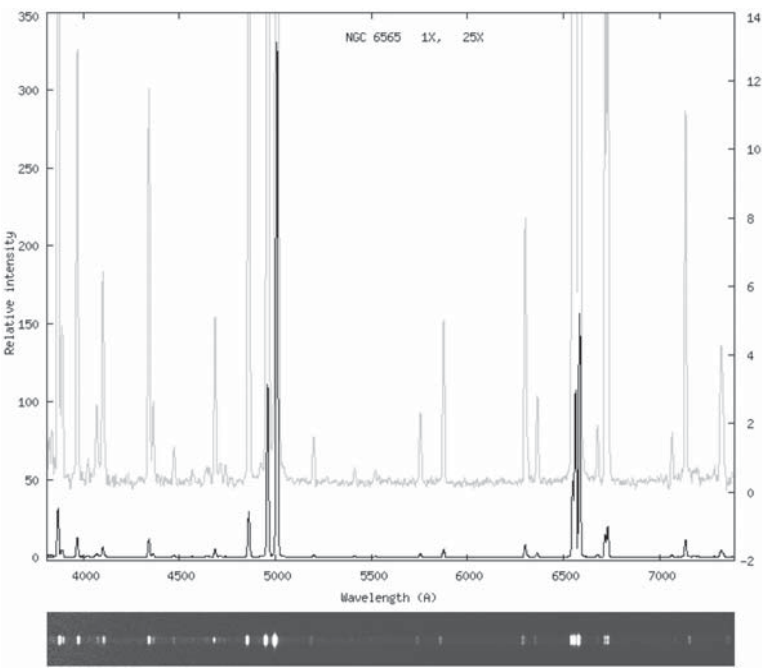

Figure 5. Spectral profile of NGC 6565

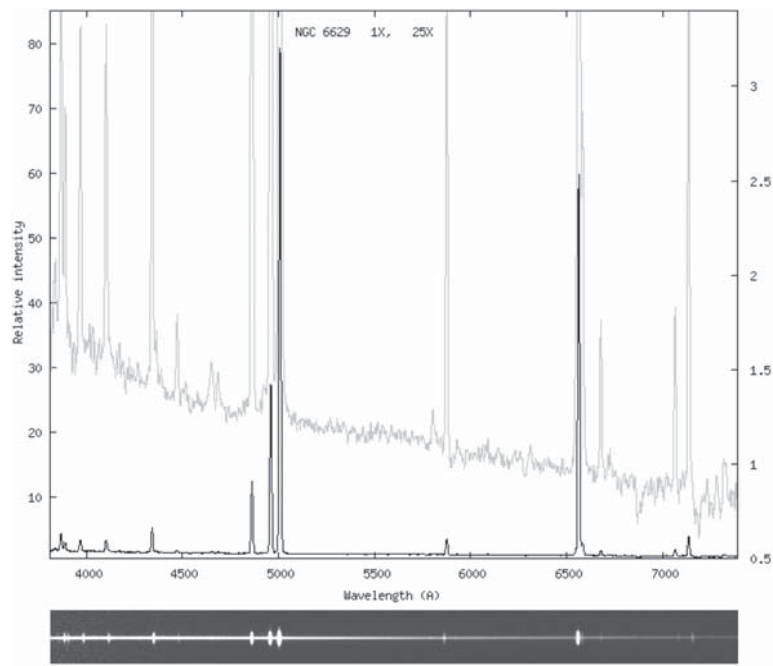

Figure 6. Spectral profile of NGC 6629

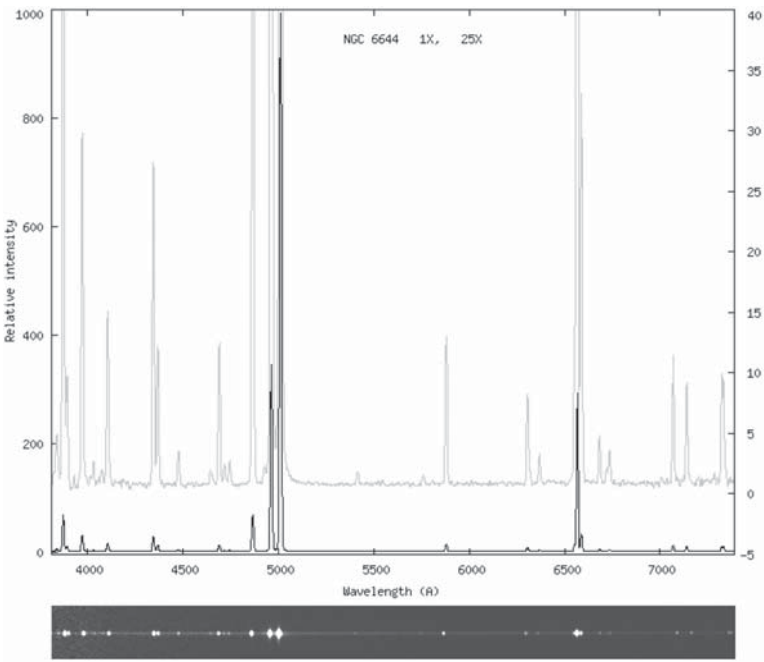

Figure 7.  Spectral profile of NGC 6644

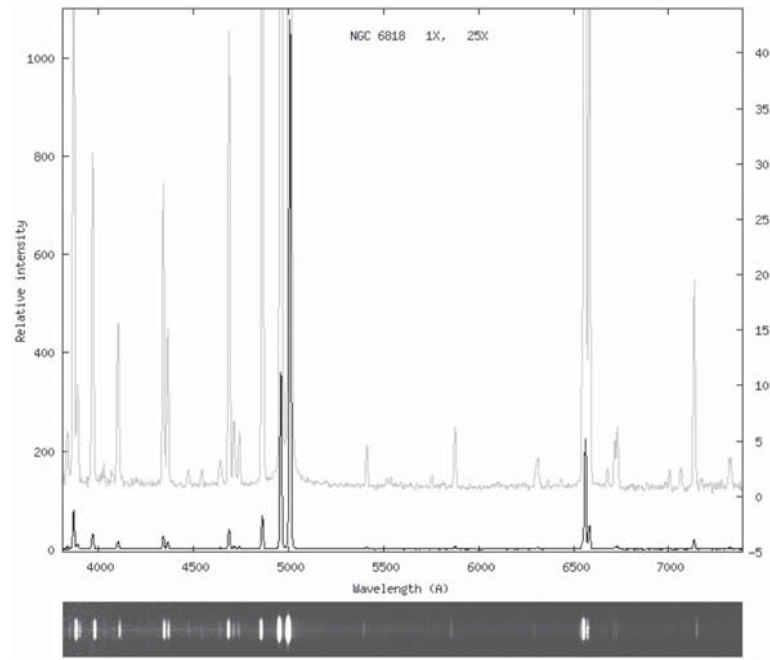

Figure 8.  Spectral profile of NGC 6818

Figures 5, 6, 7 and 8 demonstrate how changing the intensity scale of spectral profiles serves to clarify the relative line strengths of otherwise 'smaller' features, revealing the heterogeneous nature of planetary nebulae and allowing further visual comparison of excitation levels and composition. For example, the continuum in NGC 6629 in Figure 6 shows a broad increase in shorter wavelengths, perhaps a characteristic of the O-type spectral properties of its bright central star, while NGC 6818 (Figure 8) shows itself to be of a higher excitation class with its comparatively intense He II 4686Å emission. The extended profiles of NGC 6565 and NGC 6644 (Figures 5 & 7) display similar levels of excitation throughout their respective profiles, however NGC 6565 reveals itself as an 'N-rich' planetary,[12] with significantly stronger emissions in the low excitation 'red' lines of [N II] 6583Å and [S II] 6716Å, 6730Å.

## Measurement and analysis

### Line intensity, ratios and temperature

By measuring the relative line strengths, we can begin to undertake serious quantitative analysis. Line intensity is determined easily using basic measurement tools present in most spectroscopy software. In *ISIS*, a line profile and continuum tool ambiguously labelled 'FWHM' provides a variety of measured values for any selected emission or absorption feature (accomplished in most software by using the 'mouse' to define start and end points on a feature of interest). Table 2 shows the line intensities for the prominent emissions relative to Hβ for each target, derived from the values reported in the 'sum' field when using the line profile and continuum tool in *ISIS*.

It is very likely that measurement accuracy diminishes quickly for lines with low intensity. A study of NGC 6565 and NGC 6644 revealed errors as high as 50% for intensity measurements of very weak lines.[12] Even so, it is possible to derive some useful physical properties for the targets. For example, from atomic theory and assumptions based on our understanding of optically thin nebulae, the mean temperature determined by the kinetic energy of particles in the nebula can be derived using the ratio of particular line intensities.[13,14] Among the best, as it turns out, are the prominent [O III] and [N II] emissions.

Using NGC 6644 as the example, the term $R_0$ is first calculated using the measured intensity I for [O III] as follows, where λ is the wavelength:

$$R_0 = [I(\lambda 4959) + I(\lambda 5007)] / I(\lambda 4363) = (1473 + 511) / 18 = 110.2$$

An approximate electron temperature $T_e$ can be found using a simplified version of the standard formula for the [O III] ratio sourced from ref. 15: (note: coefficients vary across published electron temperature formulae as astrophysical assumptions and approximations are refined):

$$T_e = 3.3 \times 10^4 / \ln(R_0 / 8.3) \approx 12{,}800 \text{ K}$$

For strong [O III] line intensities, such as those in NGC 6644 and NGC 6818, the calculated temperatures compare well with published

Table 2. Measured line intensities scaled relative to Hβ

| Wavelength (Å) | Identity | NGC 6565 | NGC 6629 | NGC 6644 | NGC 6818 |
|---|---|---|---|---|---|
| 4340.5 | Hγ | 40 | 41 | 41 | 41 |
| 4363.2 | [O III] | 3 | 2 | 18 | 22 |
| 4685.7 | He II | 17 | – | 18 | 62 |
| 4861.4 | Hβ | 100 | 100 | 100 | 100 |
| 4958.9 | [O III] | 382 | 220 | 511 | 531 |
| 5006.8 | [O III] | 1137 | 639 | 1473 | 1587 |
| 5754.6 | [N II] | 8 | – | – | – |
| 6548.1 | [N II] | 169 | – | – | – |
| 6562.8 | Hα | 369 | 482 | 436 | 331 |
| 6583.4 | [N II] | 537 | – | 49 | 70 |
| 6716.4 | [S II] | 51 | – | – | 7 |
| 6730.7 | [S II] | 69 | – | – | 9 |

### Table 3. Measured electron temperatures vs. published values

*(Calculated electron temperatures using [O III] ratios from measured line intensities of the reduced Alpy spectra, compared to published values of the same)*

| PN | $T_e$ (K) measured | $T_e$ (K) published |
|---|---|---|
| NGC 6565 | 8,000 | 10,300[12] |
| NGC 6629 | 8,400 | 8,900[16] |
| NGC 6644 | 12,800 | 12,500[12] |
| NGC 6818 | 13,500 | 13,400[17] |

values, noting that variations of up to 'hundreds' of K appear in the literature for many planetary nebulae. Table 3 shows the calculated electron temperatures for all four targets compared with representative published values.

As an aside, the ratio derived from the line intensities of H I (the famous Balmer decrement) is often incorporated into the process to compensate for interstellar reddening. Additional line intensity ratios are similarly used to derive other physical properties of nebulae such as electron density, which is calculated using the ratio of [S II] 6716Å and 6730Å.[13]

## Radial velocity determination

The sharp spectral features of planetary nebulae would appear to make them excellent candidates for radial velocity measurements, though the literature, again, shows variations of several km/s and varying degrees of uncertainty for a selection of bright planetary nebulae. Together with *Alpy* low-resolution spectra, high-resolution spectra taken with the *Lhires III* spectrograph were also acquired on 2017 Aug 1 centred on the Hα and [O III] 5007Å emissions of NGC 6644 with total integration times of 20 and 30 minutes respectively.

Spectral images were bias, dark and flat field corrected and wavelength-calibrated using lamp spectra taken at both target wavelengths at the time of imaging. Spectra were processed in *ISIS* with heliocentric radial velocity correction enabled. The spectral resolution at 6532Å was reported as R≈ 15,000. Figure 9 shows the *ISIS*-produced high resolution spectral profile of NGC 6644 cropped and centred on the [O III] 5007Å emission. The X axis

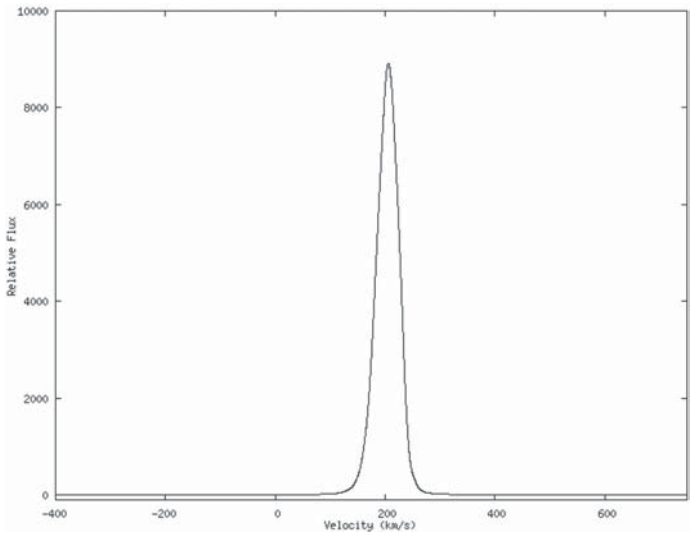

**Figure 9.** High resolution spectral velocity plot of the [O III] 5007Å emission line in NGC 6644, relative to the rest wavelength, 5006.8Å.

### Table 4. Heliocentric radial velocity measurements of NGC 6644

*(From Lhires III spectra acquired on 2017 Aug 1 measured at 3 different wavelengths)*

| Wavelength (Å) | Identity | $Vel_{hel}$ (km/s) |
|---|---|---|
| 4958.9 | [O III] | 204 |
| 5006.8 | [O III] | 202 |
| 6562.8 | Hα | 202 |

has been converted to show the emission's radial velocity shift relative to the predefined rest wavelength of 5006.8Å, a task achieved easily in software.

The line profile and continuum tool in *ISIS* is once again used, and now reports a velocity relative to the rest wavelength. In the case of NGC 6644, the reported 'position' is 202 km/s which compares well to a published heliocentric radial velocity value of 194 km/s.[18] It is notable that at this resolution, a 0.1Å shift equates to approximately 5 km/s at Hα. Despite the instrumental and measurement uncertainty, the internal velocity distributions of planetary nebulae, and the range of published values for this object, the result is pleasing. Measurements of additional lines are shown in Table 4.

Additional high resolution spectra were acquired of NGC 6565 and NGC 6629 centred on Hα on August 2 with the measured velocities presented in Table 5. The negative radial velocity of NGC 6565 indicates that it is moving towards us (effectively a blue shift of the emission lines) while the comparatively large redshift of NGC 6644's Hα line has it speeding away from us at over 190 km/s. According to one source, the comparatively high speed of NGC 6644 may be an indicator of its being a population II planetary.[12]

As an interesting test, radial velocities were also measured from the low resolution *Alpy* spectra. The line profile and continuum tool in *ISIS* was once again used, but this time the bright Hα, Hβ and [O III] 4959Å and 5007Å emission lines (and in the case of NGC 6565, [N II] 6583Å) were measured in order to calculate an average velocity. Average velocities and standard deviations for each target are presented in Table 6. Unlike the *Lhires*, the *Alpy* has not been designed with accurate velocity measurement in mind. In spite

### Table 5. Heliocentric radial velocity measurements of the Hα emission line

*(from spectra acquired on 2017 Aug 2 using the Lhires III, vs. representative published values of the same)*

| PN | $Vel_{hel}$ (km/s) measured | $Vel_{hel}$ (km/s) published[18] |
|---|---|---|
| NGC 6565 | −13 | −3.1, −4.5, −5 |
| NGC 6629 | 17 | 13, 14.6, 15 |

### Table 6. Average velocities and standard deviations

*(from the measurement of bright emission lines in Alpy spectra compared to representative published velocities)*

| PN | No. of lines measured | σ | $Vel_{hel}$ (km/s) measured | $Vel_{hel}$ (km/s) published | Δ |
|---|---|---|---|---|---|
| NGC 6565 | 5 | 10 | 12 | −4 | +16[12] |
| NGC 6629 | 4 | 11 | 37 | 14 | +23[16] |
| NGC 6644 | 4 | 16 | 226 | 194 | +32[12] |
| NGC 6818 | 4 | 14 | 11 | −13 | +24[17] |

of this, and notwithstanding the resolution difference between the two instruments, the measurements and errors appear to demonstrate the *Alpy's* potential in this area.

The impression of a systematic error manifesting itself as a positive velocity appears evident. Applying a correction by the average of these brings the measured velocities substantially closer to published values. Instrumental, calibration and other factors are potential sources for further investigation of this phenomenon. More lines per spectrum and more spectra per target would reduce the mean velocity error (the random error of the mean decreasing as the inverse square root of the number of measurements).

The use of cross-correlation measurement algorithms (also a feature in *ISIS*) should improve the accuracy by ruling out accidental errors and inconsistencies when using 'point and click' methods for selecting emission lines.

The peculiarities of radial velocity determination in planetary nebulae (the velocity differences at various points in the nebula) are notable in the literature, as are variations in published radial velocities. Despite these sources of error, the data do appear to demonstrate the degree to which even crude radial velocity measurements at low resolution can indicate general properties of these objects.

## Conclusion

This paper exemplifies the work already undertaken by many advanced amateur spectroscopists, demonstrating the power of small telescope spectroscopy. It is hoped that it provides some motivation for newcomers to embark on similar activities. The diffuse but relatively bright and compact nature of planetary nebulae makes them ideal targets for low-resolution slit spectroscopy and provides vivid examples of the properties of emission nebulae. Their strong emission lines present excellent features to experiment with and hone software measurement techniques, and the resulting analyses act as useful references as one progresses to more advanced targets.

Aside from this very practical aspect, planetary nebulae may also prove an interesting and viable area for amateur research. Despite instrumental limitations, the usual advantages of amateur astronomy apply, namely the ability to undertake long term monitoring and high cadence observation so often unavailable to our professional colleagues. Already demonstrated in other fields such as photometry, this philosophy may apply equally well to spectroscopy. The available published spectroscopic data on many bright objects including planetary nebulae are often many decades old, having been achieved using analogue instrumentation and techniques that pre-date the digital processing and software-assisted analysis now commonplace among advanced amateurs. Even with modest equipment, extensive observation can help quantify and reduce systematic errors, providing the potential to confirm, refine or even correct historical data for which only a few measurements exist.

Aside from their importance in stellar evolution, these unique and exotic objects have a rich history of observation and offer yet another interesting area for amateur exploration.

## Acknowledgments

I am grateful to the referees whose constructive comments helped to refine and improve this paper.

---

**Address:** International Centre for Radio Astronomy Research, University of Western Australia, 35 Stirling Hwy, Crawley, WA 6009, Australia. [*paul.luckas@uwa.edu.au*]